\documentclass[preprint2]{aastex}

\newcommand{\mic}{$\mu$m}

\shorttitle{Near-IR Spectral Templates Library}
\shortauthors{Winge, Riffel, Storchi-Bergmann}

\begin{document}


\title{The Gemini spectral library of near-IR late type stellar
templates and its application for velocity dispersion measurements.}


\author{Cl\'audia Winge}
\affil{Gemini Observatory, c/o Aura, Inc., Casilla 603, La Serena, Chile}
\email{cwinge@gemini.edu}

\author{Rogemar A. Riffel, and Thaisa Storchi-Bergmann}
\affil{Universidade Federal do Rio Grande do Sul, IF, CP 15051, Porto Alegre 91501-970, RS, Brazil}




\begin{abstract}

We present a spectroscopic library of late spectral type  stellar
templates in the near-IR range 2.15--2.42\mic, at R=5300--5900 resolution,
oriented to support stellar kinematics studies in external galaxies,
such as the direct determination of the masses of supermassive
black-holes in nearby active (or non-active) galaxies. The combination
of high spectral resolution and state-of-the-art instrumentation
available in 8-m class telescopes has made the analysis of
circumnuclear stellar kinematics using the near-IR CO band heads one of
the most used techniques for such studies, and this library aims to
provide the supporting datasets required by the higher spectral
resolution and larger spectral coverage currently achieved with modern
near-IR spectrographs. Examples of the application for kinematical
analysis are given for data obtained with two Gemini instruments, but
the templates can be easily adjusted for use with other near-IR
spectrographs at similar or lower resolution. The example datasets are
also used to revisit the ``template mismatch'' effect and the
dependence of the velocity dispersion values obtained from the fitting
process with the characteristics of the stellar templates. The
library is available in electronic form from the Gemini web pages at
\url{http://www.gemini.edu/sciops/instruments/nearir-resources/?q=node/10167}

\end{abstract}





\section{Introduction}

Spectral templates, usually late type stars,  are required for the
analysis of kinematical data on external galaxies or other stellar
ensembles
\citep[e.g.][]{emsellem01,marquez03,barbosa06,ganda06,cappellari07,dumas07,riffel08,cappellari09,riffel09}.
In the near-infrared (hereafter near-IR), the most commonly used features are the CO
overtone bands at $\lambda > 2.29$\mic. These band heads originate
from evolved stars and are the strongest features in the 1--3\mic\
spectral range of stellar systems. The features are deep and sharp,
and at least the first two overtones are located in regions of the IR
spectrum relatively clean from telluric lines.

Although observational and theoretical libraries exist at lower
spectral resolutions \citep[$R\leq 3000$,
e.g.][]{kleinmann86,wallace97,ramirez97,forster00,ivanov04}, no
comprehensive set of stellar kinematic templates was available to be used with the $R\sim 6000$
configuration of the two Gemini NIR instruments used for stellar population
kinematic studies in external galaxies - the Near-infrared
Integral Field Spectrograph (NIFS) and the Gemini Near-Infrared
Spectrograph (GNIRS) with the 111 l/mm grating  (both long-slit and Integral
Field Unit - IFU), and, therefore, all observing programs using those
configurations would invariably spend some science time taking a small
set of stellar spectra to use as templates. This led to a constant
multiplication of data taking, since those targets were defined as
program calibrations and were not made available to other users until the end
of the default 18-months proprietary period.

During semester 2006B at Gemini South, given the unusually poor
conditions over the whole semester, a Director's Discretionary ``poor
weather''  program was specifically carried out to provide the NIR
community with a larger set of late (F7 to M3 types) stellar spectra
in the range 2.24--2.43\mic, including the four CO overtone bands, at
$R \sim 5900$ resolution.  Most of the targets were also observed at a
slightly bluer spectral range (2.15--2.33\mic) to improve usefulness
for NIFS users, overlapping with the red setting on the first two CO
bands.

To the original sample of 29 stars observed with GNIRS, another 11
were added from NIFS observations obtained as part of programs
GN-2006A-SV-123, GN-2006B-Q-107, and GN-2007A-Q-25, covering the full
range 2.1--2.5\mic\ at a similar resolution to that of the GNIRS
data. 

\section{Observations and Data Reduction}

\subsection{The GNIRS data}

GNIRS \citep{elias98} is a multi-configuration spectrograph originally
deployed at Gemini South, now moved to Gemini North, that allows for
several combinations of resolution and wavelength coverage. The two
modes most commonly used for the science relevant here were
the long-slit with the 32 l/mm or 111 l/mm gratings, which yield
resolutions of up to R$\sim$2000 and R$\sim$6000, respectively
(depending on the slit used); and the Integral Field Unit with the
same gratings, where the internal optics projected each
0.15$^{\prime\prime}$ slice into 2 detector pixels in the dispersion
direction, resulting a single fixed resolution similar to the one
obtained with the 0.3$^{\prime\prime}$ long-slit, and a field of view
of $3.2^{\prime\prime} \times 4.8^{\prime\prime}$. The IFU has been
decommissioned at Gemini North, given the capability overlap with
NIFS.
 
The observed sample was selected from a list kindly provided by
G. Doppmann, compiled from the literature \citep[mostly based
in][]{strobel97}, and the selection was based exclusively on
observability: targets which were visible for as long as possible
during the semester, bright enough to provide the desired S/N on a
reasonable on-source time under cloudy, poor seeing conditions, and
having a hot (A0--A7) star close enough -- and bright enough -- to be
used for telluric correction. In addition, both target and telluric
stars had to have a bright ($V < 13$mag) star available in the guide
probe patrol field.

The observations were done using the IFU
\citep{allington06,allington07} with the grating 111 l/mm, yielding a
resolving power of R$\sim$ 5900.  The list of observations is given in
Table~\ref{gnirsobs}, where for each object are listed the $V$
magnitude, effective temperature and surface gravity when available,
as well as the observing dates and the spectral resolution, measured
from the arc lines, for the ``blue'' (centered at 2.245\mic)  and
``red'' (centered at 2.335\mic) settings, respectively.

The observing conditions also determined the instrument configuration:
to achieve $R=5900$ with GNIRS in long slit mode, one would have to use
the $0.30^{\prime\prime}$ slit, implying in very large slit losses
under poor seeing conditions (FWHM $> 0.80^{\prime\prime}$\ in
K). Given the superior GNIRS IFU performance in the K band (over 90\%
of that of the equivalent long slit mode), there was only a small loss
in sensitivity by using the IFU+111 l/mm grating configuration.

The standard group of observations included a science target, one
telluric standard star, a set of calibrations comprising
three arcs and a set of ten QH lamp flats. Calibrations (arcs and flats)
were usually observed right after the science target, or after a set
of targets was observed, but before the grating was moved to another
configuration.

Observing sequences were defined as several (2 to 5) repeats of ABBA
sequences, with a 4$^{\prime\prime}$\ offset between the A (object) and B (sky)
positions in a direction perpendicular to the  long axis of the IFU
field-of-view, large enough to move a centered object completely
off to sky.  On-target efficiency with this setup gets reduced by 50\%,
but it avoids the problem of overlapping PSF wings due to the small
size of the IFU if trying to dither on source.

The GNIRS data frames as delivered to the Gemini Science Archive are
in the standard Gemini MEF (Multi-Extension Format), where the primary
header unit (PHU, extension [0]) includes all header information from
telescope, environmental monitoring system and instrument; and the
data extension [1] contains the pixel values. Data reduction was
performed using the tasks in the {\em gemini.gnirs} {\sc iraf}
package\footnote{The Gemini Data Reduction package can be obtained
through the Gemini web pages at
\url{http://www.gemini.edu/sciops/data-and-results?q=node/10795}.},
release Version 1.9, of July 28, 2006, and comprised the following steps:

For the calibrations (flats and arcs):

\begin {enumerate}

\item {\em nsprepare}: this task reformats the data files to add the
IFU Mask Definition File (MDF), which contains the information on the
position of the individual slices in detector coordinates,  and
applies the linearity correction to the pixel data. The resulting file
contains the PHU, one binary table extension with the MDF, and one data
extension with the actual pixel values.

\item {\em nsreduce}: cuts out each of the 21 IFU slices according
to the MDF inserted above to a separate data extension. No dark
correction was applied to either flats or arcs. 

\item {\em nsflat}: combines the ten frames by data extension, using {\em
ccdclipping} for rejection  and normalizing by the median of the
illuminated area in each slice, as defined in the MDF. On average, the
processing resulted in a S/N $\sim$\ 200-300 for each extension, with
exception of slices 1 and 21 (which were partly vignetted) and slice
13 (which was damaged).

\item used {\em gemcombine} to average the three processed arc frames
to improve visibility of faint lines, then {\em nswavelength} to
obtain the wavelength solution from the combined arc. For the standard
Gemini Ar lamp, there were four lines in the ``red'' setting, and
six in the ``blue'' setting. A low order polynomial
(legendre order=3) was used for the fitting, with residuals of the
order of 0.15\AA\ or less.

\end{enumerate}

For the science data and telluric stars:  

\begin{enumerate}

\item {\em nsprepare}: same as for the calibrations.  

\item {\em nsreduce}: after cutting the individual slices to separate
data extensions, the task corrects the frames for the flat-field, and
subtracts adjacent object-sky frames.

\item {\em nsstack}: since we had only one position with actual data
(the B position was blank sky), and the targets were all bright  point
sources observed under poor seeing conditions, we simply stacked all A
positions without any effort to improve alignment of the individual
frames by shifting according to the offsets registered in the
headers. In most cases all frames were within $0.3^{\prime\prime}$
tolerance, but there were a few observations where drifts of up to
$0.8^{\prime\prime}$ were seen, usually due to clouds or very poor
seeing affecting the guiding performance.

\item {\em nstransform}: the wavelength transformation is applied to the stacked frame 

\item {\em nsextract}: finally, we interactively extracted the
spectrum from each slice, in order to exclude those with very low
signal, since the targets were not always well centered in the IFU
field-of-view, the two edge slices, and the damaged slice when the
spectrum happened to fall within the damaged region. The output from
this task is still a MEF file, with each data extension containing a
1D spectrum. 

\item  used a simple {\sc iraf} cl script wrapped around {\em
specred.scombine} to combine all valid spectra obtained in the
previous step. At this point, a single 1D standard FITS spectrum is
created, but most of the information contained in the PHU of the MEF
files is lost, as {\em scombine} propagates the header of the first
extension included in the combining list, ignoring the content of the
PHU. 

\item finally, applied the telluric correction in the science data,
using the standard {\em specred.telluric} task, and combined all
spectra for those targets observed in different nights.

\item added back the header information lost in step 6, corresponding
to the content of the PHU of the MEF frame obtained in
step 5. This preserves all telescope, instrument and program
information.

\item removed the continuum shape by fitting a low
order polynomial to the final 1D spectrum, and corrected all  spectra to
rest velocity, by measuring the central wavelength of a strong,
isolated line (Mg{\sc i} at 2.2814\mic) and using it as a reference
zero point for all datasets. This corrected in a single step the
intrinsic radial velocity and any zero-point offset that could exist
from the wavelength calibration.

\end{enumerate}

Figures~\ref{gnirsplot} to \ref{gnirsplotlast} show the resulting
spectra, with ``red'' and ``blue'' settings combined when it applies. 

\subsection{The NIFS data}

NIFS \citep{mcgregor03} is an IFU spectrograph with coronographic
masks, optimized for use with the Gemini North Adaptive Optics (AO)
system Altair. In the K band, the standard NIFS configuration yields a
2-pix spectral resolution of $R=5290$ (at 2.2\mic), covering the entire
1.99--2.40\mic\ range in a single setting.

The data were obtained either as program calibrations for
GN-2006A-SV-123 and GN-2007A-Q-25, and therefore based solely in
observability and brightness; or as part of a ``poor weather''
program GN-2006B-Q-107, and in this case following the same
rationale as the GNIRS sample (bright enough for poor conditions, with
proper telluric and guide stars available). The stars observed as part
of 06A-SV-123 and 07A-Q-25 used the AO wavefront sensor for guiding, while for
06B-Q-107, the AO fold was parked and guiding was done using the
peripheral wavefront sensor only (no AO correction).

The selected instrument configuration was the K\_G5605 grating and
HK\_G0603 filter, resulting in a FWHM for the arc lamp lines of
~3.2\AA. Each observations consisted of five individual exposures,
with the star centered on the array then offset to each
corner. Table~\ref{nifsobs} lists the observations, where for each
object we present the spectral type, observing date(s), spectrum
central wavelength (in \mic), and spectral resolution from the arc
lines (in \AA). 

The data reduction was done in a similar way as for the GNIRS data,
using the corresponding tasks in the {\em gemini.nifs} {\sc iraf}
package. The reduction procedure included trimming of the images,
flat-fielding, sky subtraction, wavelength and s-distortion
calibrations. Removal of the telluric bands and flux calibration was
executed in single step by interpolating a black body function to the
spectrum of the telluric standard star. Finally, the continuum shape
was removed from the spectrum of each star (using the {\sc iraf} task
{\em continuum}), normalizing the continuum fluxes to unity.  

Similar to the GNIRS spectra, the extraction procedure results in loss
of the primary header content. This was added back to the spectra in
order to propagate the relevant instrument/telescope information. The
spectra were also shifted to a common reference point using the same
Mg{\sc i} line. Figures~\ref{nifsplot} and \ref{nifsplotlast} show the
resulting NIFS spectra.

\section{Template fitting examples.}

Assuming that a galaxy spectrum is the convolution of a composite
stellar spectrum (combination of all the individual stars
present in the local stellar population) with the line-of-sight
velocity distribution, the most commonly method used to extract the
velocity dispersion from the galaxy spectrum employs a
cross-correlation technique using one or more stellar spectra as
templates \citep{tonry79}. This technique, however, is sensitive to
the effect of template mismatch
\citep[e.g.][]{silge03,emsellem04,riffel08}, with the fitting results
being affected by the characteristics of the individual stellar
spectrum used as template. To obtain a reliable measurement, one
should provide the fitting algorithm with a variety of template
stellar spectra, which will be internally attributed different fitting
weights to obtain the best result.

\citet{silge03} (hereafter SG03) have examined the issue and concluded
that it is the equivalent width (EW) -- or alternatively, the shape -- of the CO band head
in the templates that affects the fitting, not the details of the
spectral type. Their analysis, using a subsample of stars from the
library of \citet{wallace97} (herafter WH97), showed a trend for the
velocity dispersion measured for the galaxy to {\it increase} as the
equivalent width of the template star's band head {\it increases}.
They therefore argue that the template sample should span a range of CO
equivalent widths, more than simply of spectral types, as the EW of the CO bands is a function of the effective
temperature and surface gravity of the star, increasing
with decreasing surface gravity or effective temperature.

Figure~\ref{libew} shows the EW of the first CO overtone ($^{12}$CO
(2-0) at 2.294\mic) in our sample, measured using the {\sc iraf} task
{\em onedspec.splot} in the window 2.293--2.322\mic, that is, from
the blue edge of the (2-0) band head to the blue edge of the (3-1)
band head, plotted as a function of the effective temperature (T$_{eff}$). 
This window is larger than the one used by SG03 (2.288--2.305\mic),
but avoids trying to define a ``continuum'' point among the resolved
$^{12}$CO (2-0) resonance lines. The resulting EW are about twice the
value obtained using SG03 wavelength range. The EW
in the data ranges from less than 5 to over 30\AA. For the template
stars not in the sample of \citet{strobel97}, the value of T$_{eff}$\ was
obtained from Table~15.7 of Allen's Astrophysical Quantities, 4th
Edition \citep{allen00}, using a simple linear interpolation between the
two nearest spectral types when necessary. Figure~\ref{libco}
presents a sample of spectra in the spectral range including the
Mg{\sc i} 2.2814\mic\ line (left), the first CO and part of the second ($^{12}$CO (3-1) at 2.323\mic) overtones, to illustrate the different
equivalent widths and profiles.

As an example of the use of the templates, we selected actual data
from three Gemini programs, GS-2005B-Q-65, a GNIRS/IFU+111~l/mm
observation of the Seyfert galaxy NGC7582, centered at 2.24\mic\
\citep{riffel09}; GN-2006A-SV-123, and GN-2007A-Q-25, NIFS observations of
the Seyfert galaxies NGC4051, and NGC4258, respectively
\citep{riffel08,riffel10}.  The analysis technique used was
the penalized pixel fitting (pPXF) method of \citet{cappellari04},
which allows to use one or several template stellar spectra, and to
vary the weights of the contribution of the different templates to
obtain the best fit. A complete description of the data, reduction
procedure and analysis is presented in the first two references above.

Figures \ref{n4051map} and \ref{n7582map} present full 2D velocity
dispersion maps for NGC4051 and NGC7582 respectively, using a single
star with smaller (HD105028) and larger (HD30354) CO equivalent
widths, and the full library. Sample spectra and corresponding fits,
extracted from the data cubes, are show in Fig.~\ref{galfit}. The
effect is quite evident: the large-scale structures in the maps are
similar for all stars, but the structures in smaller scales and the mean
$\sigma$ values vary significantly, with {\it larger} equivalent widths in
the templates resulting in {\it lower} velocity dispersion values measured
from the galaxy spectra. 

This is the opposite result as found by SG03 and prompted a more
detailed analysis of the assumed dependency of the fitting result with
the stellar template characteristics. To start with, we repeated the
fitting of the NGC4051 off-nucleus spectrum show in
Fig.~\ref{galfit}, using each individual template star in the present
library. The galaxy spectrum was then degraded in resolution and the
fitting process repeated using each of the stars in the WH97 library
with detectable CO absorption (63 objects in total). Since the
velocity dispersion of NGC4051 is quite low, and the resolution of the
WH97  templates is R$\sim$3000, we repeated the same process for the
nuclear spectrum of NGC4258, which has $\sigma \sim$170 km s$^{-1}$,
closer to that of the object analyzed by SG03. The results are shown
in Figure~\ref{sigmaew}. The fitting results using our templates are
shown as filled squares; the open triangles are the results from the
fitting using the stars from WH97. An average error of 8 km s$^{-1}$
is assumed (estimated by \citet{riffel08} from the fitting of the
NGC4051 datacube). The dotted lines indicate the value of $\sigma$
obtained by fitting the galaxies using all stars from our library
simultaneously. 

The top panel, with the results for the NGC4258 nuclear spectrum, give
an indication of why we see a different trend from SG03. Those authors
selected only 9 stars from the WH97 library, and the dependence
found with EW rests strongly in the $\sigma$\ value obtained from the
template at the lower end of the EW range (see their Figure~4,
remembering that the EW plotted here are about a factor of 2 larger
due to the different window used in the measurement). The scattering of
points for the stars with EW $<$ 10\AA\ would indicate that, had they
chosen another star, they would have perhaps found a very weak to no
dependence of $\sigma$ with EW, or even a negative correlation as the
NGC4051 fitting seems to indicate. 

For this later galaxy, the low dispersion velocity coupled with the
lower resolution of the WH97 create a quite distinct negative
correlation, with $\sigma$ {\it decreasing} as the EW
increases. Looking at the results using our higher resolution
templates, this correlation is much weaker, and essentially rests --
just as before -- in the one or two templates at the lower end of the
EW range. Therefore, the only systematic effect we can derive from our
plots is that for templates with lower EW to overestimate the $\sigma$
in the galaxy spectrum when such velocity dispersion is already low
and close to the limits allowed by the spectral resolution of the data
and templates. 

We are, therefore, led to conclude that the resulting velocity
dispersion is not directly dependent on the EW of the CO band in the
template star, just as SG03 concluded that it was not directly
dependent on its spectral type. For NGC4258, the two K0 III stars in our library
result on $\sigma$\ values more than 60 km s$^{-1}$ apart, but at the same
time it can be seen in the top panel of Fig.~\ref{sigmaew} that
templates with essentially the same EW of around 11\AA\ can yield results
ranging from 140 to 240 km s$^{-1}$.

It is more likely that there is not a simple dependence of the fitting
result with any particular characteristic of the template star
spectrum, but rather that each template, taken in isolation, yields a
result that is better or worse, when comparing with the actual
$\sigma$ of the galaxy, depending on how close that individual template is
to the overall characteristics of the stellar population contained in
the galaxy spectrum. This example illustrates the importance of
using a complete stellar library in the determination of the velocity
dispersion in galaxies, just as one is needed when analyzing stellar
population ages or metallicities in composite spectra.

Figure~\ref{correlation} shows the effect of the slight difference in
resolution between the NIFS and GNIRS data. The left panel presents
the values of $\sigma$ measured from all the spectra in the NGC4051
datacube using only the GNIRS stars as templates against the same
measurements using only the NIFS stars. There is an excellent
correlation between the two values, but with a systematic offset to
lower $\sigma$ values using the GNIRS templates. If we consider only
the nominal resolving power of the respective gratings, there is a
$\sim$ 5.9 km s$^{-1}$ difference in resolution between the two
instruments at 2.3\mic. The right panel presents the values obtained
for $\sigma$ after degrading the resolution of the GNIRS templates to
the nominal resolving power of the NIFS grating. The 8 km$^{-1}$ error bars in the
top left corner correspond to the composite error derived by
\citet{riffel08}. The residual tendency for the NIFS data to
overestimate the value of $\sigma$ at low values can be traced to the
uneven distribution of template characteristics of the two samples (the
NIFS templates, being fewer and concentrated towards K5 III types or
later, do not by themselves provide a full enough
representation of the stellar population of the galaxy).

The individual template spectra also show a range of spectral
resolution as measured from the FWHM of the arc lines. The resolution
of the NIFS templates ranges from 3.12 to 3.26\AA, while that of the
GNIRS ones ranges from 2.71 to 3.16\AA. When using the templates to
fit NIFS and GNIRS data, the difference is small enough that this
effect gets mixed with the noise and fitting errors, but  users that require the best precision
should convolve each individual spectrum in the library to match the
resolution of the data being fitted. Tables \ref{gnirsobs} and
\ref{nifsobs} present the resolution of each stellar template, as
measured from the arc lines.

One of the most important results that can be obtained from deriving
stellar velocity dispersions in the inner regions of Active Galactic
Nuclei is the mass of the central supermassive black hole
($M_{BH}$). The  ($M_{BH}$) vs $\sigma$ correlation
\citep{gebhardt00, ferrarese00, tremaine02} can be represented as
$\log(M_{BH}/M_\odot) = \alpha + \beta\ log(\sigma_*/\sigma_0)$, where
$\alpha = 8.13 \pm 0.06$, $\beta= 4.02 \pm 0.32$  and  $\sigma_0 =
200$ km s$^{-1}$.  If we take the measurements presented in
Fig. \ref{n4051map}, the value obtained for the bulge central velocity
dispersion of the galaxy NGC4051 using the full library  for the
fitting is $\sim$60 km s$^{-1}$, which from the above equation results
in $M_{BH} \sim 1.1\times10^6 M_\odot$. On the other hand, if only the
star HD105028 (a K0III, the spectral type most commonly used as
template) was available for this analysis, the result would be a
velocity dispersion of $\sim$105 km s$^{-1}$, and thus $M_{BH} \sim
1\times 10^7 M_\odot$, a full order of magnitude difference!

It is not directly obtainable from the analysis done here if
the effect of using the same single stellar template to determine the
velocity dispersion in a sample of objects will be a systematic over-
or under-evaluation of $\sigma$. It  may be an effect proportional to the
actual value of the velocity dispersion, or even a pure scattering of
the points around the actual values, depending on how close a
representation of the local stellar population is the chosen stellar template.
In any case, if the mass of the supermassive black hole for a certain
object or sample of objects is determined by other methods, while  the
value of $\sigma$ is measured by template fitting, the effect of template
mismatch can introduce further scattering on the data points in the
$M_{BH}$) -- $\sigma$ correlation.

It is, in any case, quite evident that the improvements in the quality
of the datasets and in the analysis methods have led studies of
stellar velocity dispersion fitting in external galaxies to join the
ranks of the more general issue of composite stellar population
analysis, in the sense that a well-known, as complete as possible,
template base is needed.

\section{Availability of the library}

The Gemini library of late type stellar templates  can be downloaded
from the Gemini website as individual objects or in two compressed
files for the NIFS and GNIRS stars: 

\url{http://www.gemini.edu/sciops/instruments/nearir-resources/?q=node/10167}

The page contains links to the final processed spectra and details of the
data reduction. It also presents the history and release
notes for the different versions of the library. The version described
in this paper corresponds to V1.5 from 2009 Jan 31. 

The data are presented in standard FITS format, and the user can
select either the GNIRS ``red'' (2.24--2.42\mic) or ``blue'' (2.15--2.32\mic)
spectral ranges at their native spectral binning; or the combined
spectrum (when both ranges were observed), rebinned to 1\AA/pix. The
NIFS spectra are also presented at native binning and rebinned to
1\AA/pix. All data were kept with their original spectral resolution,
so the users can more adequately adjust the templates to the actual
resolution of their own datasets.

The community at large is welcome to download all or part of the
library as needed, and users of GNIRS and NIFS are encouraged to
explore its use as an alternative to requesting further observation of
spectral standards with their science programs. All the GNIRS data
collected under program GS-2006B-DD-3 has been made public from 
the start in the Gemini Science Archive. The NIFS
data were subject to the standard proprietary period. If the raw data are
re-processed to be used in papers or publications, please use the
standard Gemini acknowledgment text for archival data and the
appropriate program IDs.



\acknowledgments

The authors would like to thank the Gemini Deputy Director and Head of
Science, Jean-Rene Roy, and the former Gemini South Head of Science
Operations, Michael West, for the support and time allocation for our
GNIRS program. Many thanks as well to all the Gemini South observers
and SSAs that so positively believed that no conditions were ever too
poor to give GS-2006B-DD-3 a chance!

We thank the anonymous referee for pointing out the issue that led to
the more thorough discussion of the velocity dispersion dependency with
template characteristics, which significantly enhanced the scope of
this paper. 

Based on observations obtained at the Gemini Observatory, which is
operated by the Association of Universities for Research in Astronomy,
Inc., under a cooperative agreement with the NSF on behalf of the
Gemini partnership: the National Science Foundation (United States),
the Science and Technology Facilities Council (United Kingdom), the
National Research Council (Canada), CONICYT (Chile), the Australian
Research Council (Australia), Minist\'erio da Ciencia e Tecnologia
(Brazil) and Ministerio de Ciencia, Tecnolog\'{\i}a e Innovaci\'on Productiva
(Argentina). Data taken under program IDs: GS-2006B-DD-3,
GN-2006A-SV-123, GN-2006B-Q-107, and GN-2007A-Q-25.



{\it Facilities:} \facility{Gemini:South}, \facility{Gemini:Gillett}.

\clearpage

\onecolumn


\begin{figure}
\figurenum{1a}
\includegraphics[angle=180,scale=.90]{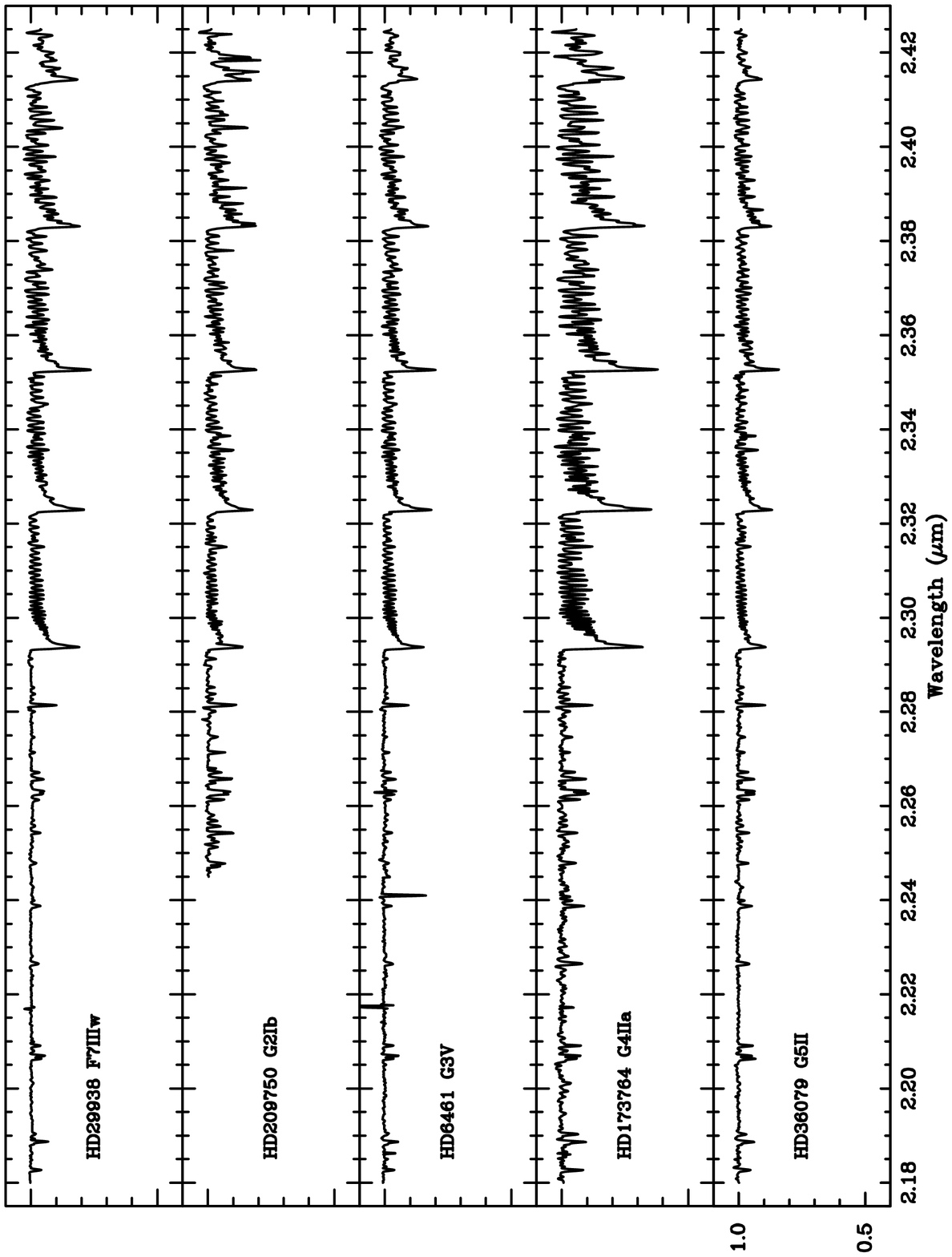}
\caption{The GNIRS template spectra, arranged by spectral type. \label{gnirsplot}}
\end{figure}


\begin{figure}
\figurenum{1b}
\includegraphics[angle=180,scale=.90]{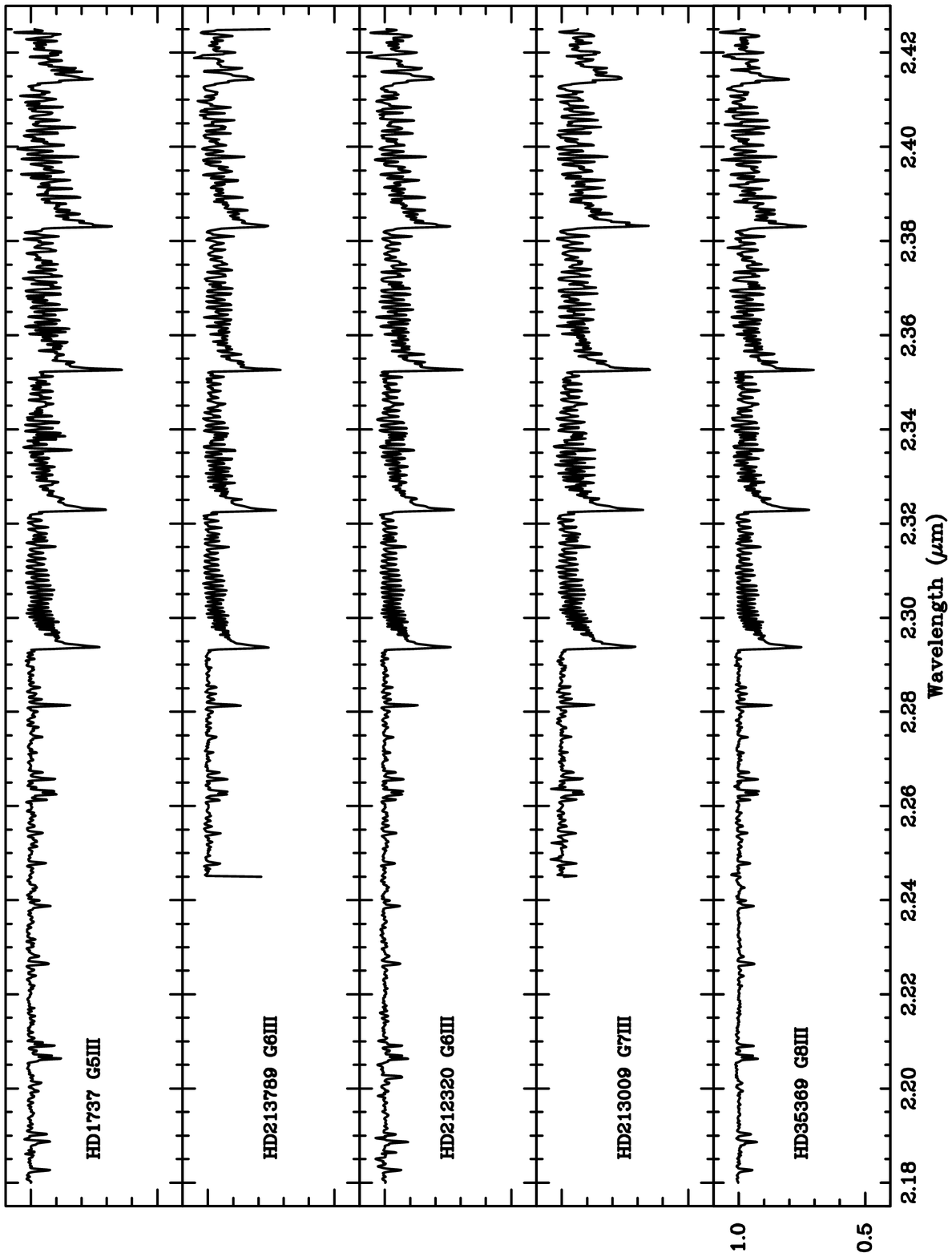}
\caption{Continued.}
\end{figure}


\begin{figure}
\figurenum{1c}
\includegraphics[angle=180,scale=.90]{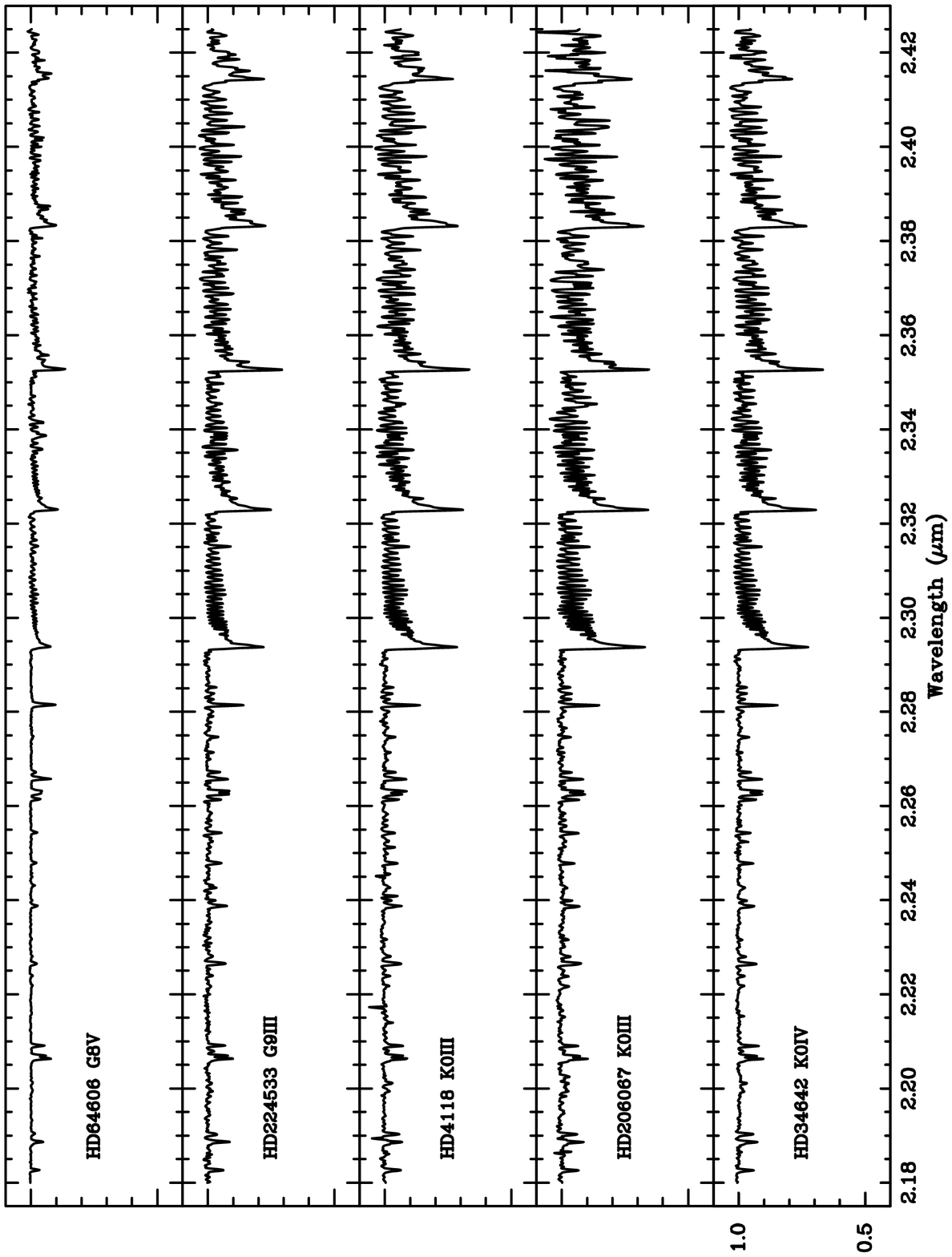}
\caption{Continued.}
\end{figure}


\begin{figure}
\figurenum{1d}
\includegraphics[angle=180,scale=.90]{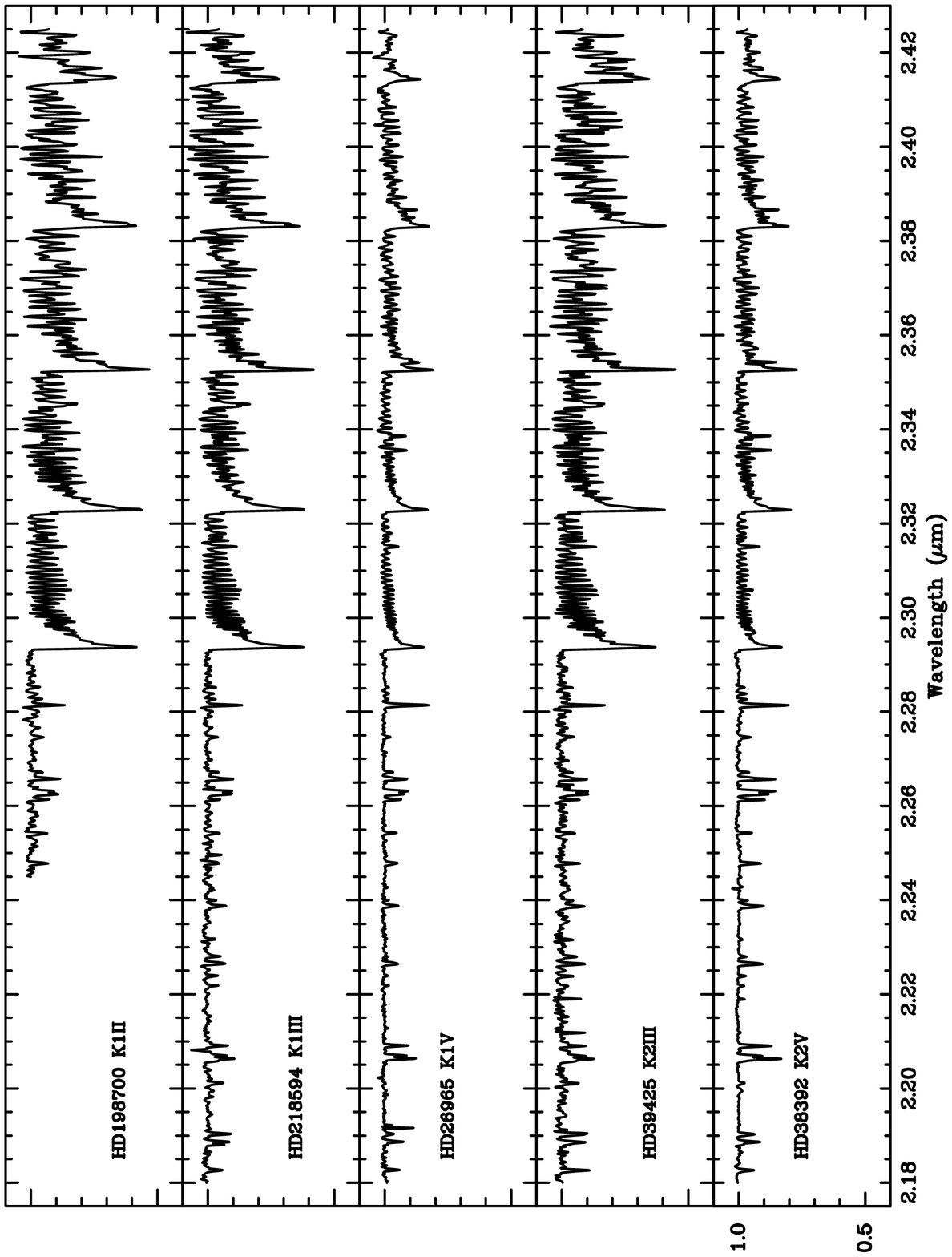}
\caption{Continued.}
\end{figure}


\begin{figure}
\figurenum{1f}
\includegraphics[angle=180,scale=.90]{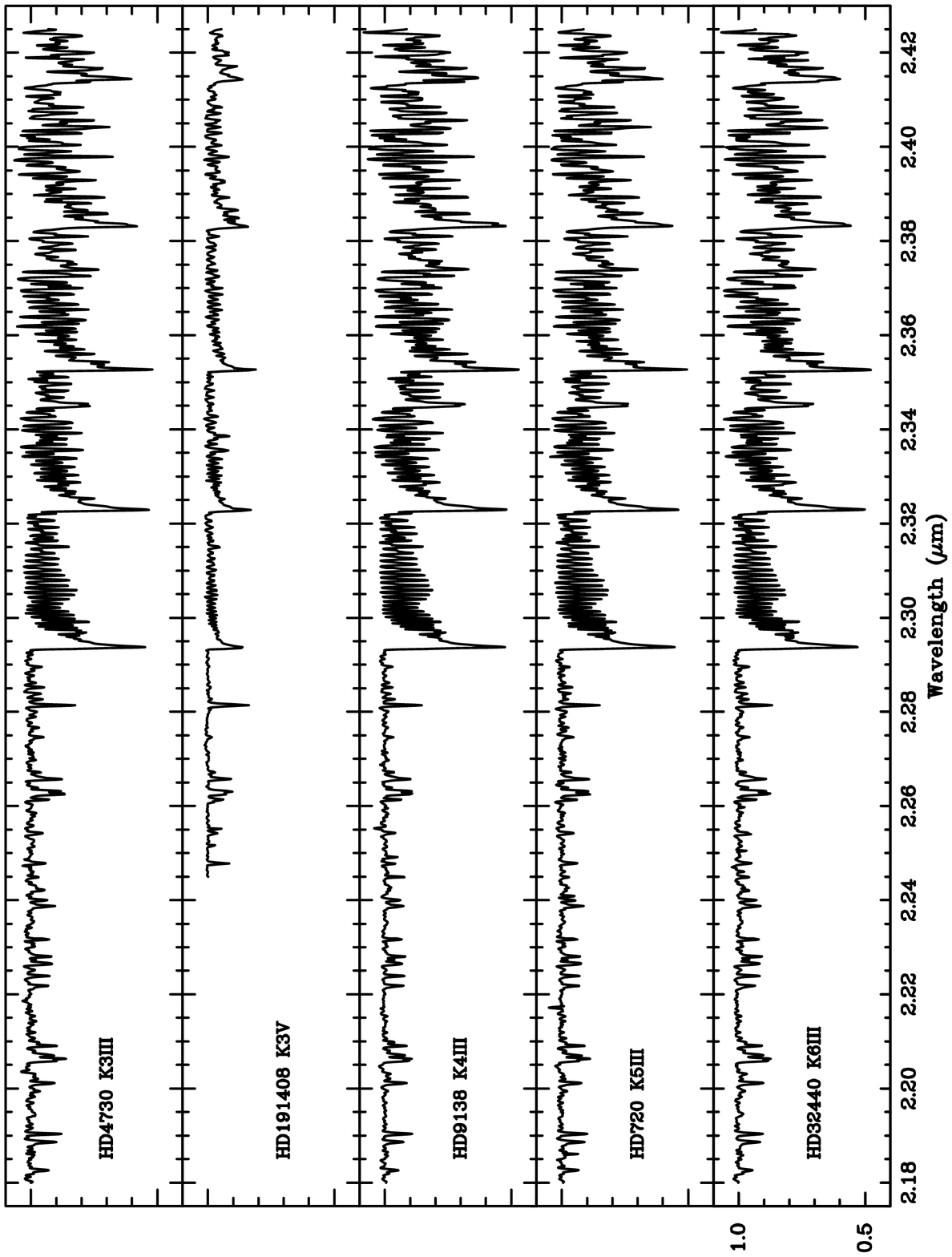}
\caption{Continued.}
\end{figure}


\begin{figure}
\figurenum{1g}
\includegraphics[angle=180,scale=.90]{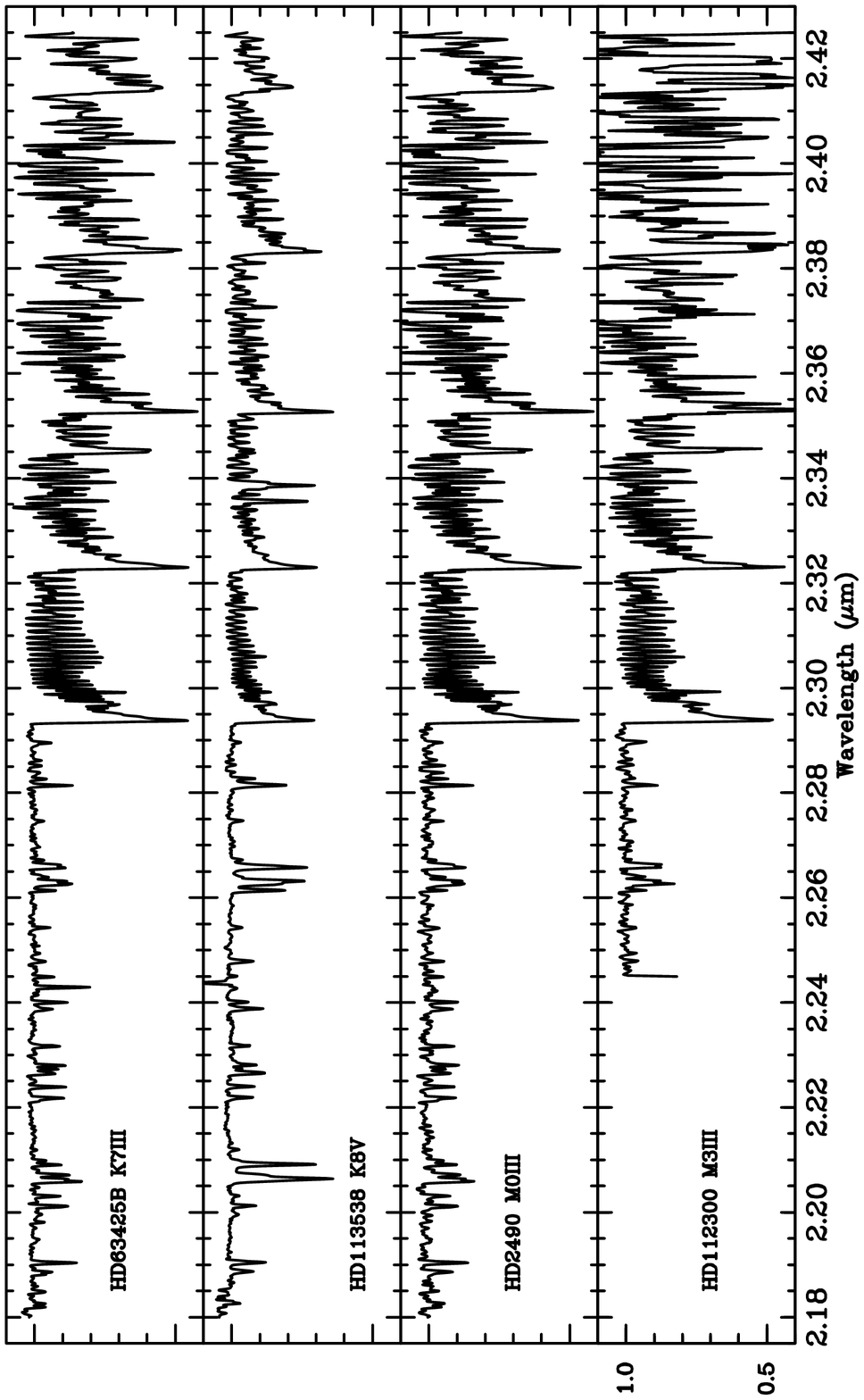}
\caption{Continued.\label{gnirsplotlast}}
\end{figure}


\begin{figure}
\figurenum{2a}
\includegraphics[angle=180,scale=.90]{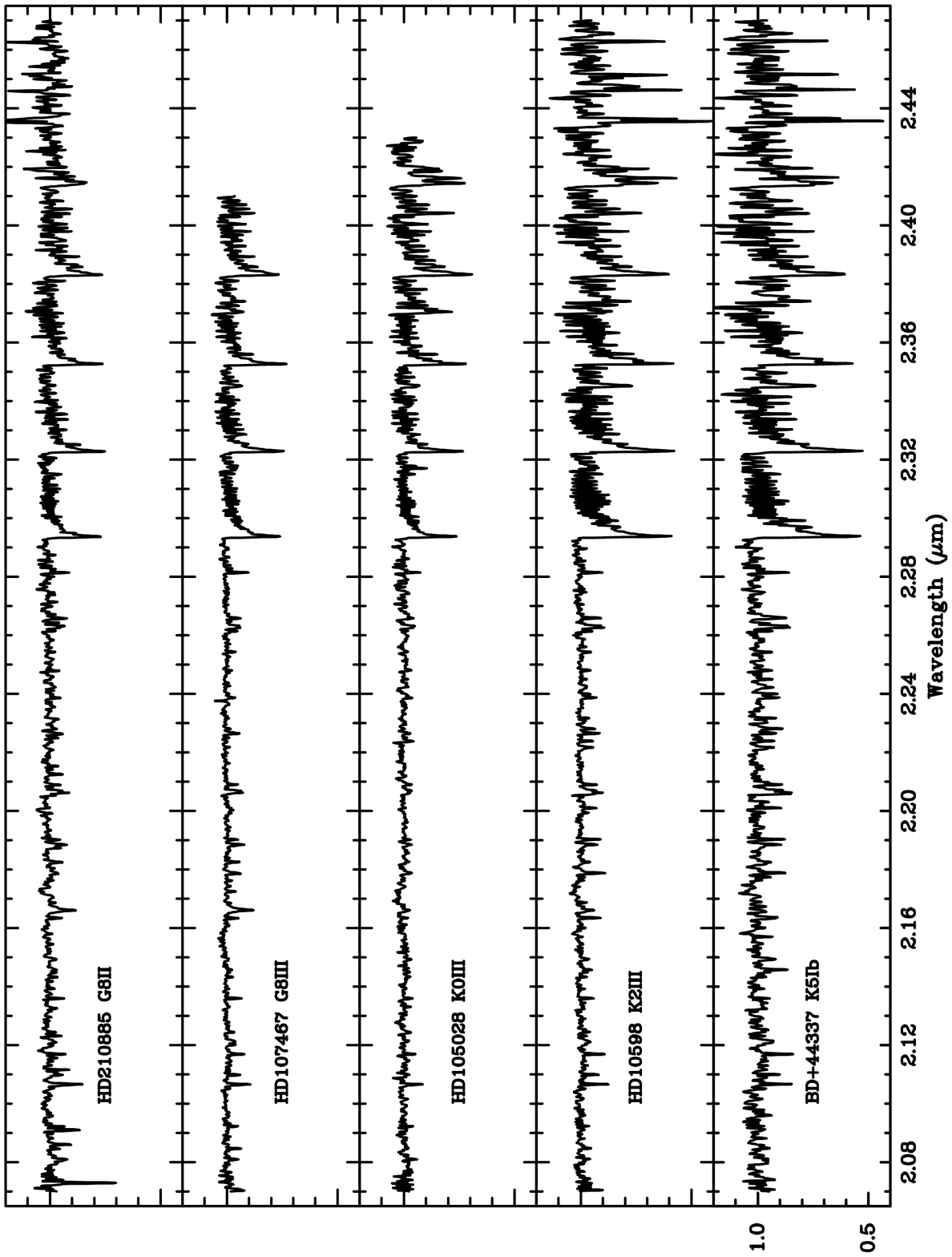}
\caption{The NIFS template spectra, arranged by spectral type.\label{nifsplot}}
\end{figure}


\begin{figure}
\figurenum{2b}
\includegraphics[angle=180,scale=.90]{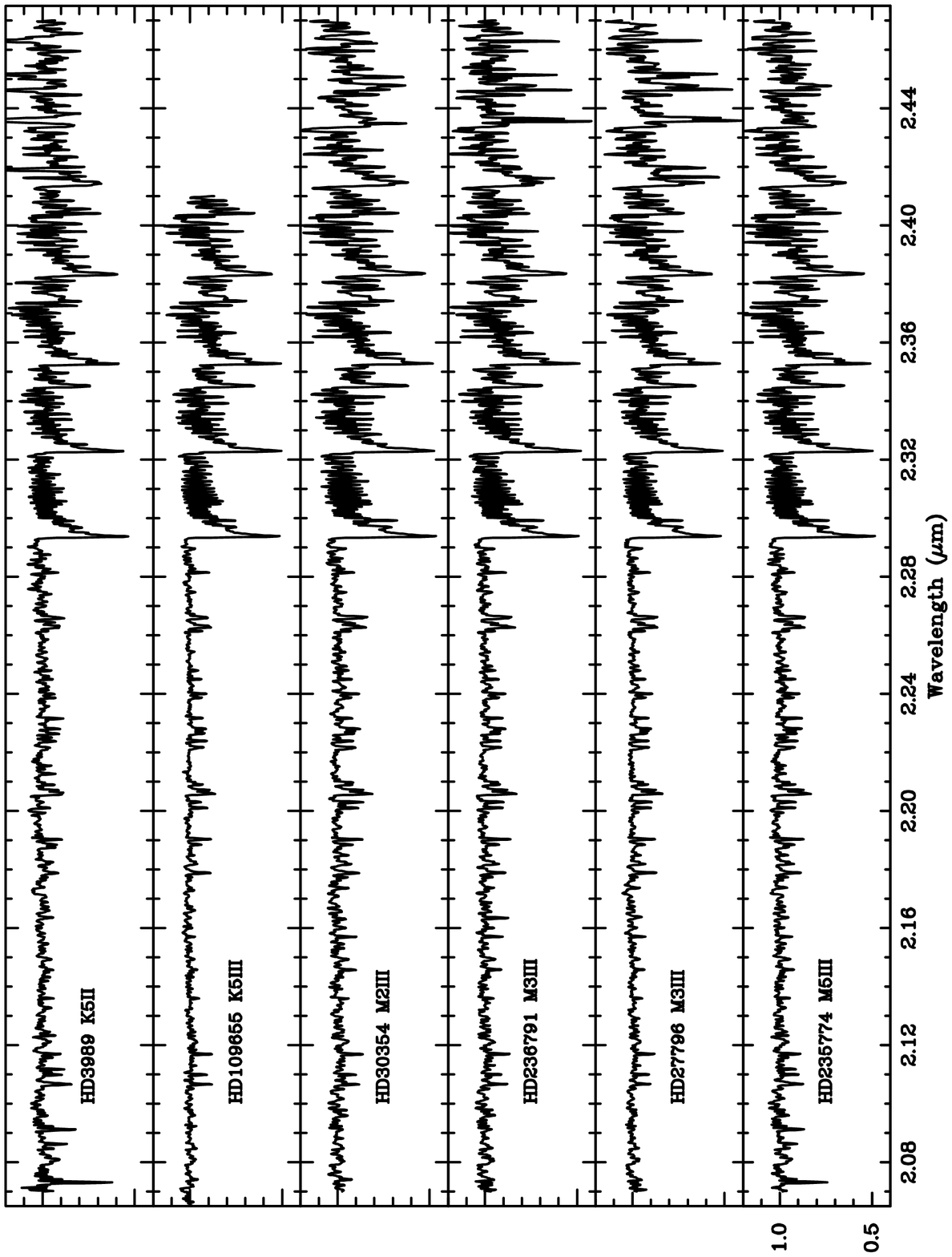}
\caption{Continued. \label{nifsplotlast}}
\end{figure}


\setcounter{figure}{2}

\begin{figure}
\includegraphics[angle=90,scale=0.60]{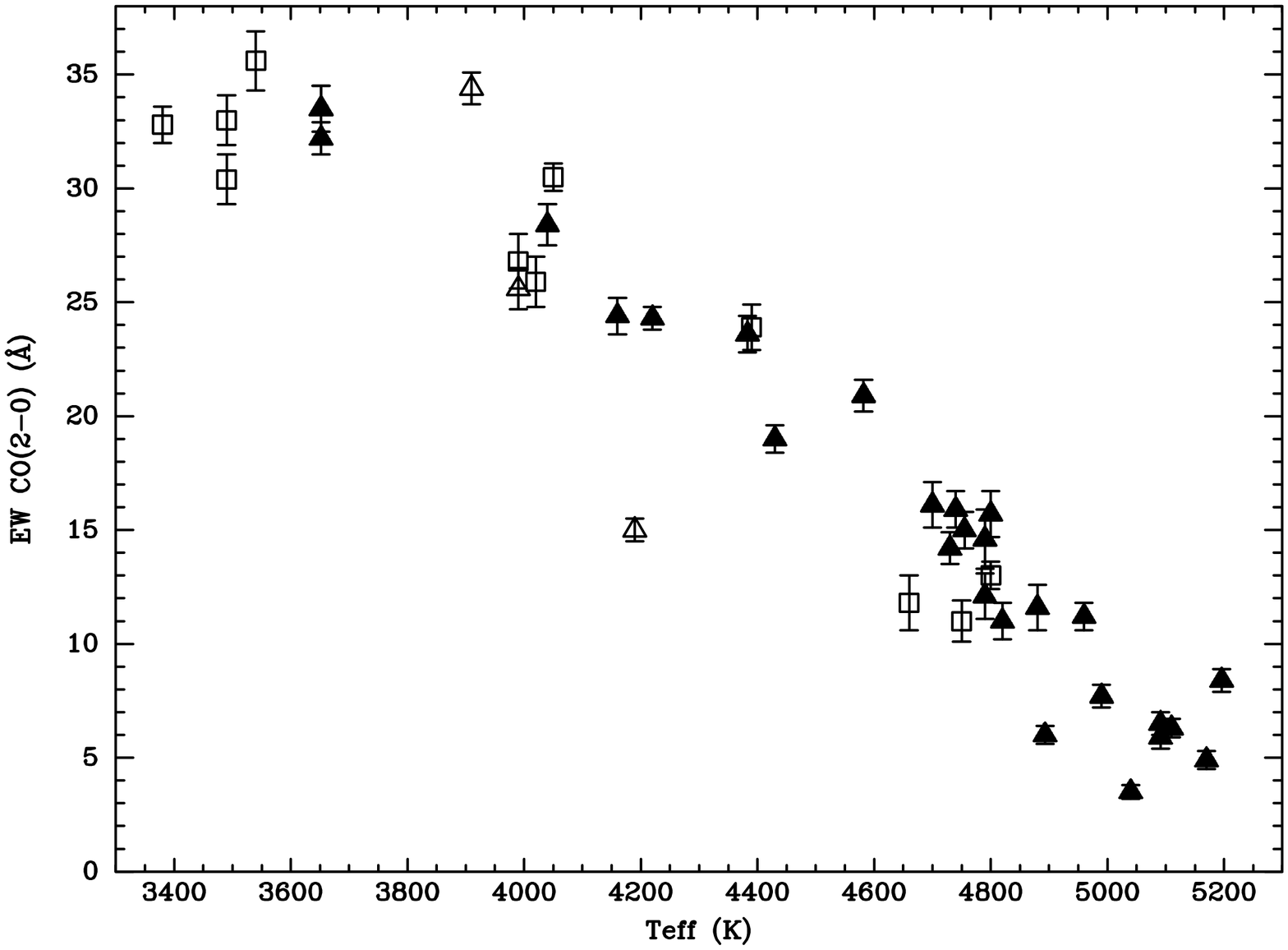}
\caption{Equivalent width of the CO (2-0) 2.294\mic\ band for our
sample, plotted as a function of  effective temperature. The EW has
been measured in the window 2.293--2.322\mic\ and the errors
correspond to upper and lower placement of the continuum. The GNIRS
templates are represented as triangles, the NIFS sample as
squares. Closed symbols are stars with temperatures from
\citet{strobel97}, while open symbols correspond to stars for which
the temperature was approximated from the spectral
type. The star with EW=15\AA\ and T$_{eff}\sim$4200K is HD113538 (see
note in Table~\ref{gnirsobs}). \label{libew}}
\end{figure}


\begin{figure}
\includegraphics[angle=-90,scale=0.60]{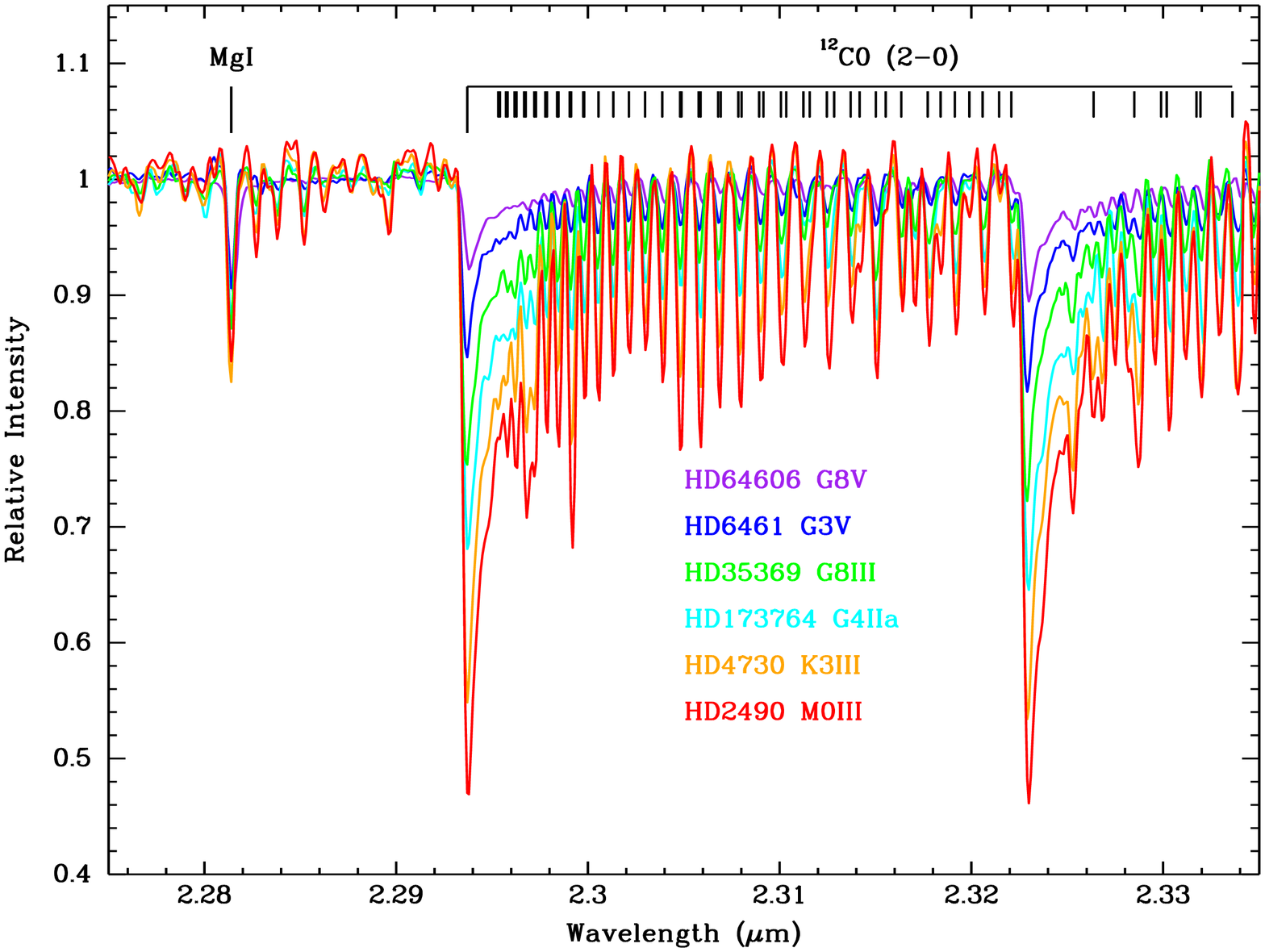}
\caption{Sample GNIRS template spectra including the Mg{\sc I}, the first
and part of the second CO overtones, to illustrate the different
equivalent widths and profiles. Note how the individual $^{12}$CO
(2-0) resonance lines are resolved. The CO line identification was
taken from the high resolution atlas of
\citet{hinkle95}. \label{libco}}
\end{figure}


\begin{figure}
\epsscale{1.0}
\plotone{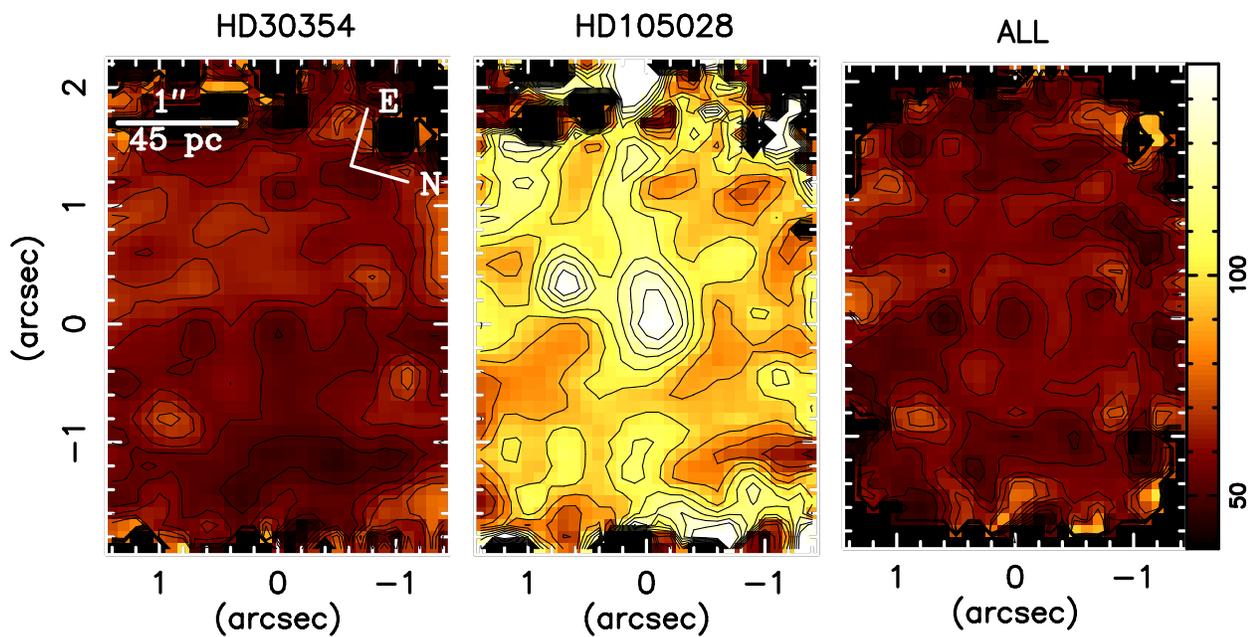}
\caption{Velocity dispersion maps obtained from NIFS observations of
NGC4051, using as spectral templates for the fitting: (left) only one
star, HD30354 (EW(CO) = 35.6\AA); (middle) only one star, HD105028
(EW(CO) = 11.8\AA); (right)  the full set of templates contained in
this library. Not only the derived velocity dispersion values, but
also some of the structures in small spatial scales are different in
each case.  \label{n4051map}}
\end{figure}


\begin{figure}
\epsscale{.50}
\plotone{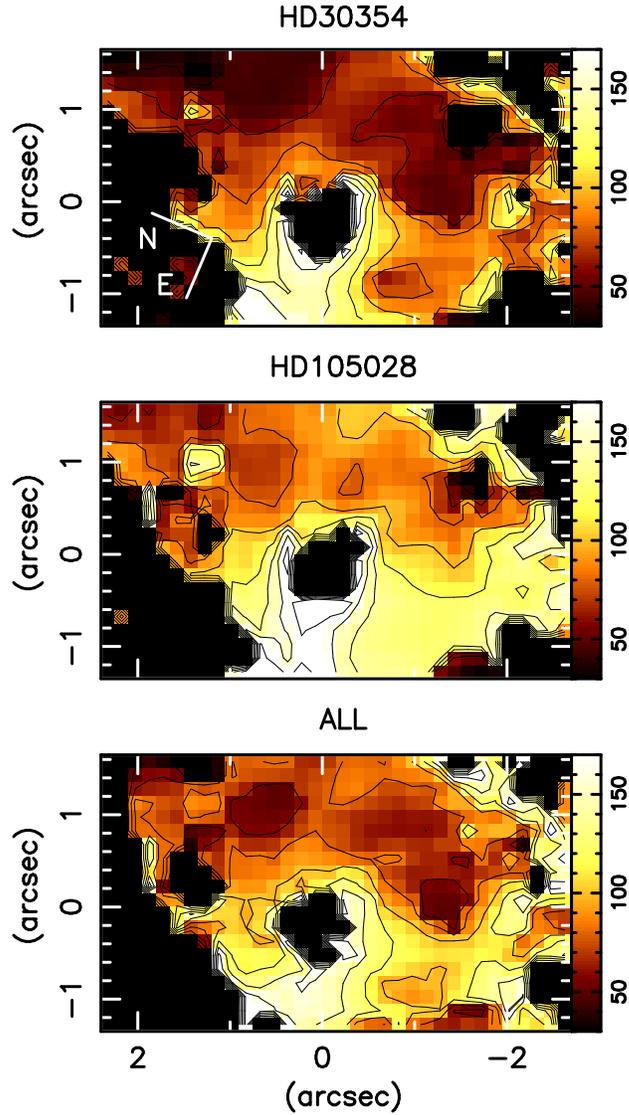}
\caption{Velocity dispersion maps obtained from GNIRS observations of
NGC7582, using as spectral templates for the fitting: (top) only one
star, HD30354 (EW(CO) = 35.6\AA); (middle) only one star, HD105028
(EW(CO) = 11.8\AA); (bottom) the full set of templates contained in
this library. In this case, the overall structure of the three maps is
more similar, but the effect of template mismatch in the resulting
values for $\sigma$ is quite evident. \label{n7582map}}
\end{figure}


\begin{figure}
\epsscale{1.0}
\plottwo{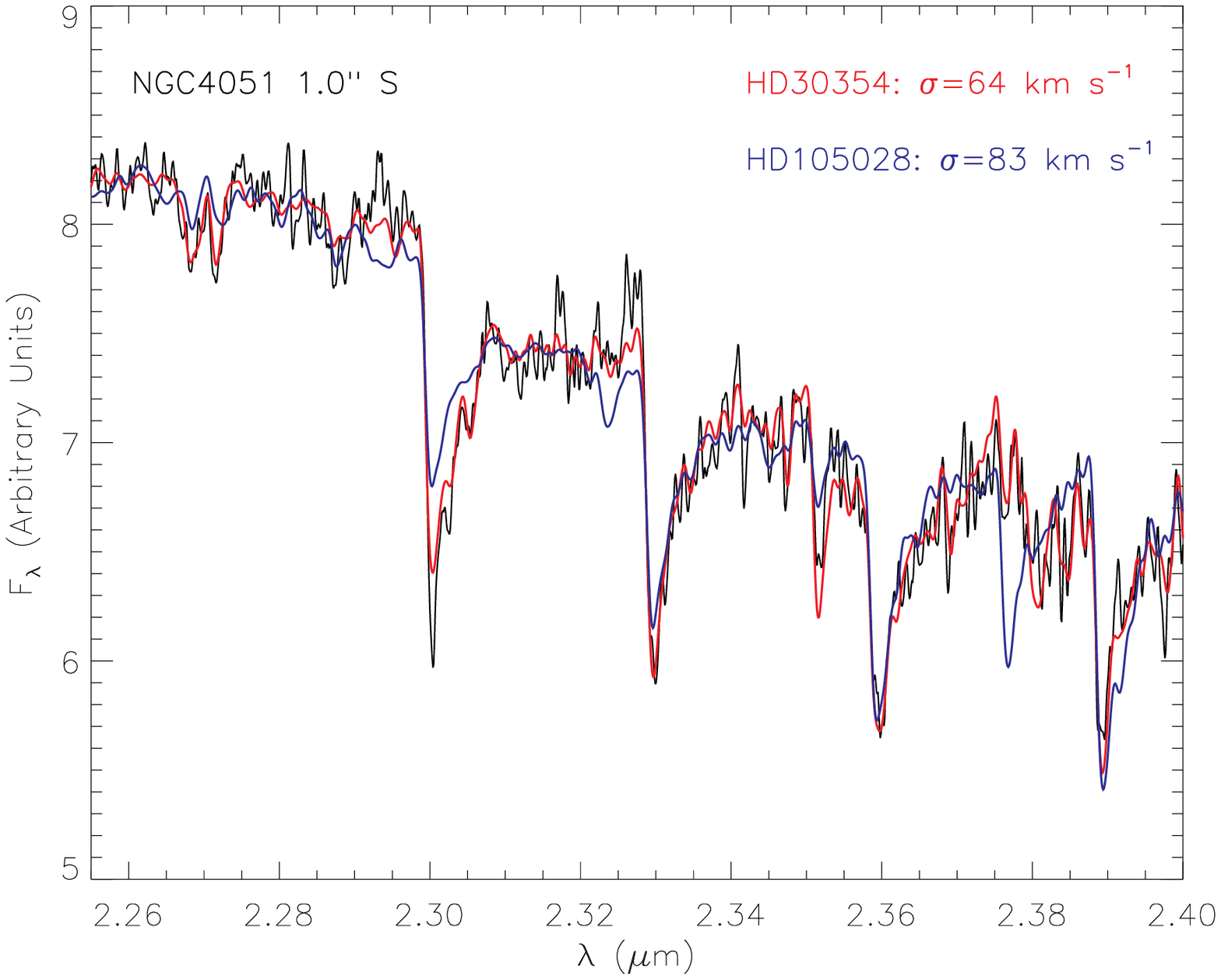}{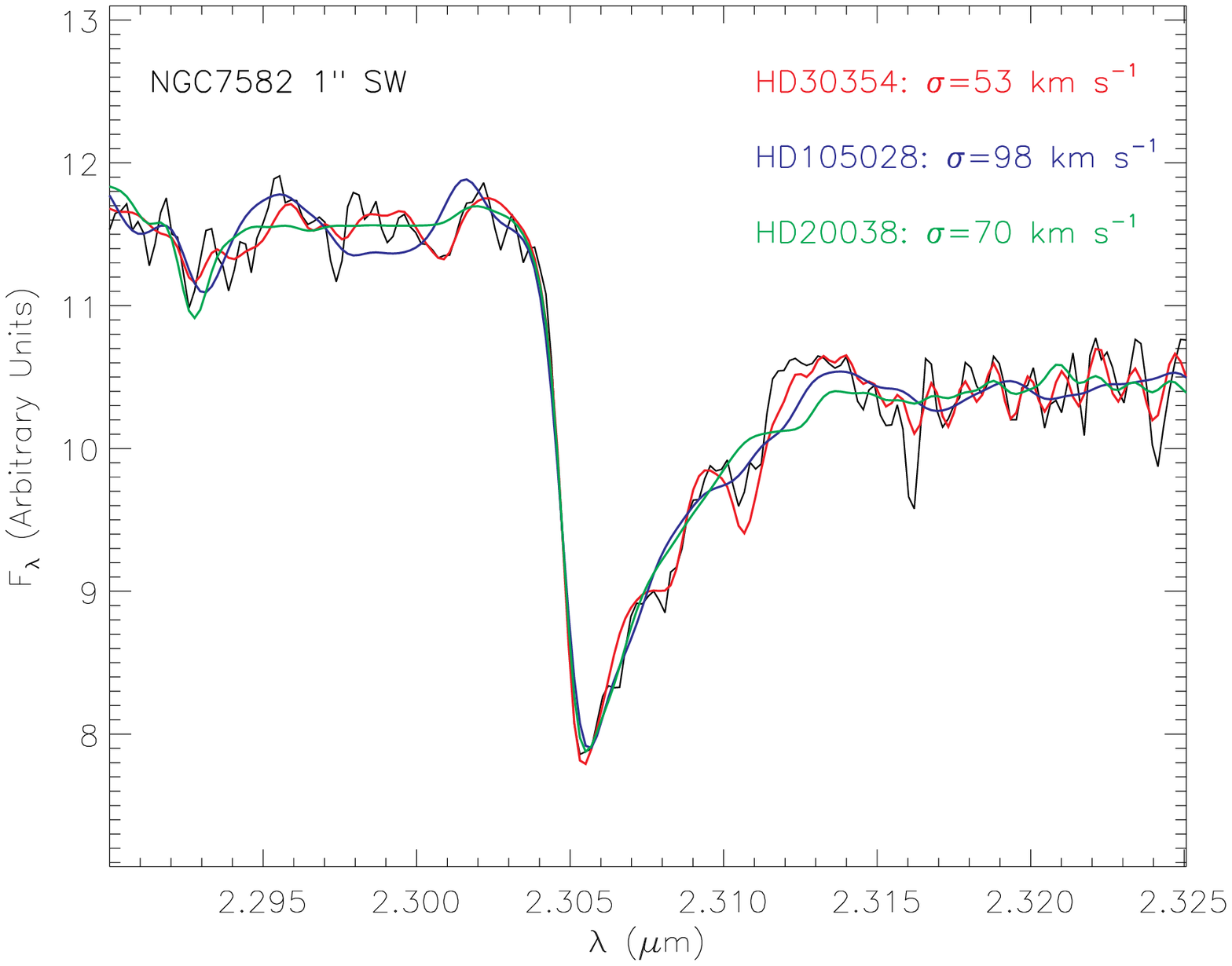}
\caption{Example of the template fitting results  for spectra
extracted from the data cubes shown in Fig. \ref{n4051map}\ and
\ref{n7582map}. (left) Spectrum of a region located
1$^{\prime\prime}$\  South of the nucleus of NGC4051 (black), with the
convolved HD30354 (red) and HD105028 (blue) templates overploted. The
resulting value of $\sigma$\ for each fit is shown in the top right
corner. (right) Same as previous, for the spectrum of a region located
1$^{\prime\prime}$\ South-West of the nucleus of NGC7582, with the
addition of the fit using HD20038 (EW(CO)=8.4\AA) as template (green). \label{galfit}}
\end{figure}


\begin{figure}
\includegraphics[angle=90,scale=0.60]{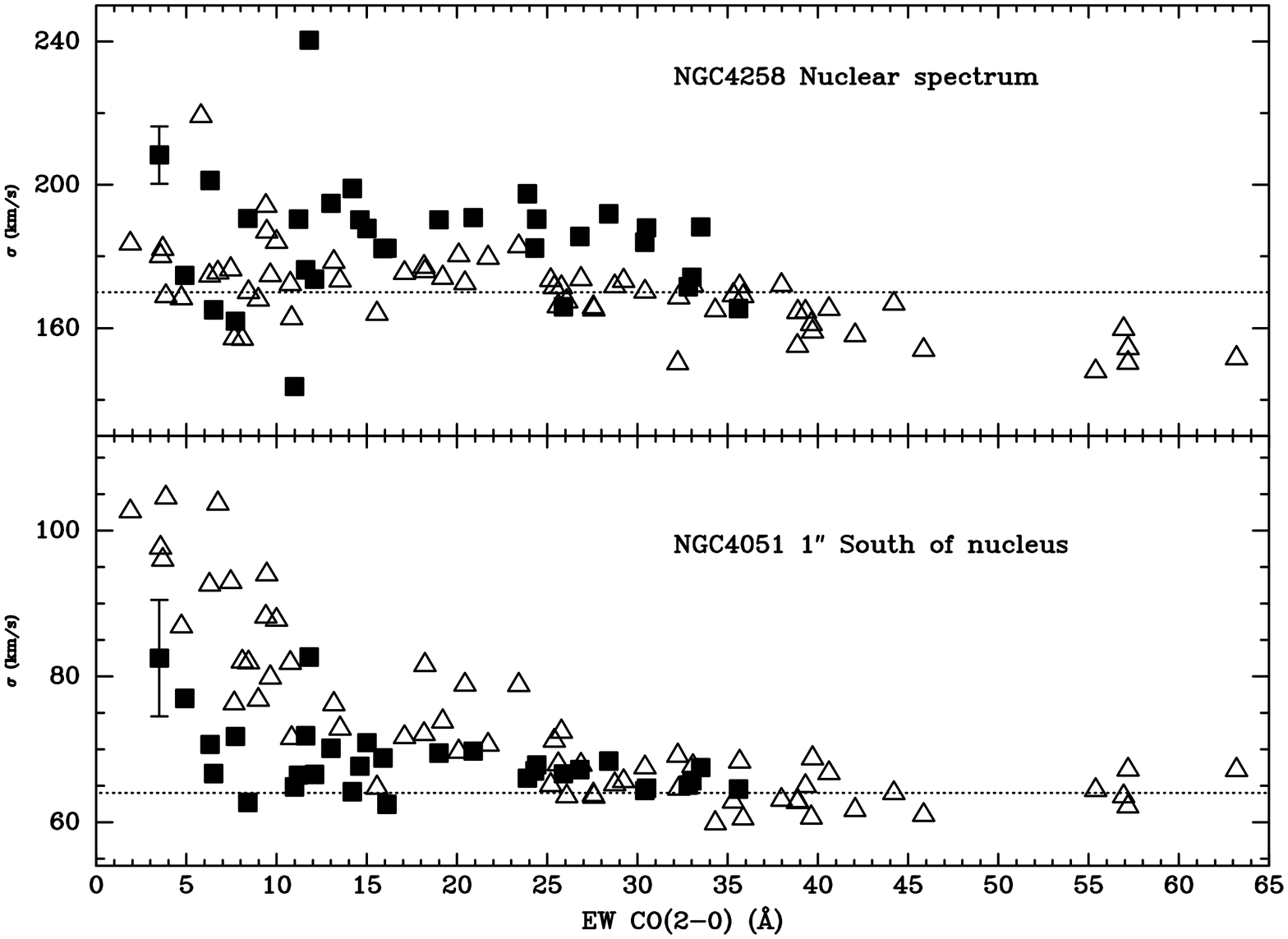}
\caption{Velocity dispersion resulting from fitting individual template stars from this library (filled
squares) and that of \citet{wallace97} (open triangles), plotted as a
function of the equivalent width of the $^{12}$CO (2-0) absorption
band. The top panel presents the results for the nuclear spectrum of
NGC4258; the bottom panel for an off-nuclear spectrum in NGC4051. The
dashed line is the velocity dispersion value found by using all the
stars in the present library as initial input to the fitting
program. The error bar corresponds to an average error of 8 km
s$^{-1}$; individual errors tend to be slightly larger (10-12 km
s$^{-1}$) for low EW templates fitting a low $\sigma$\ spectrum such as
NGC4051, or slightly smaller for larger template EW and $\sigma$. \label{sigmaew}}
\end{figure}


\begin{figure}
\epsscale{1.0}
\plottwo{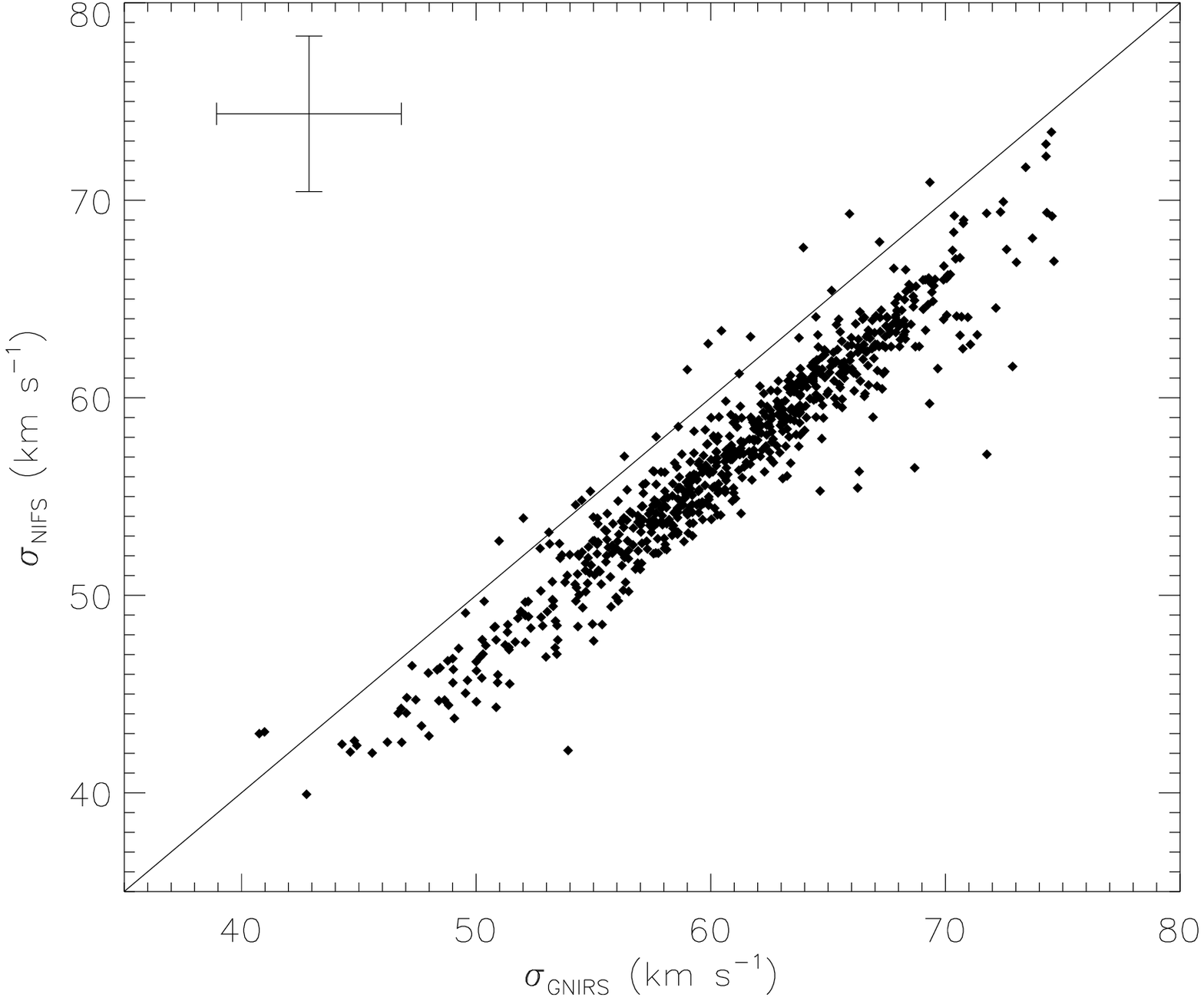}{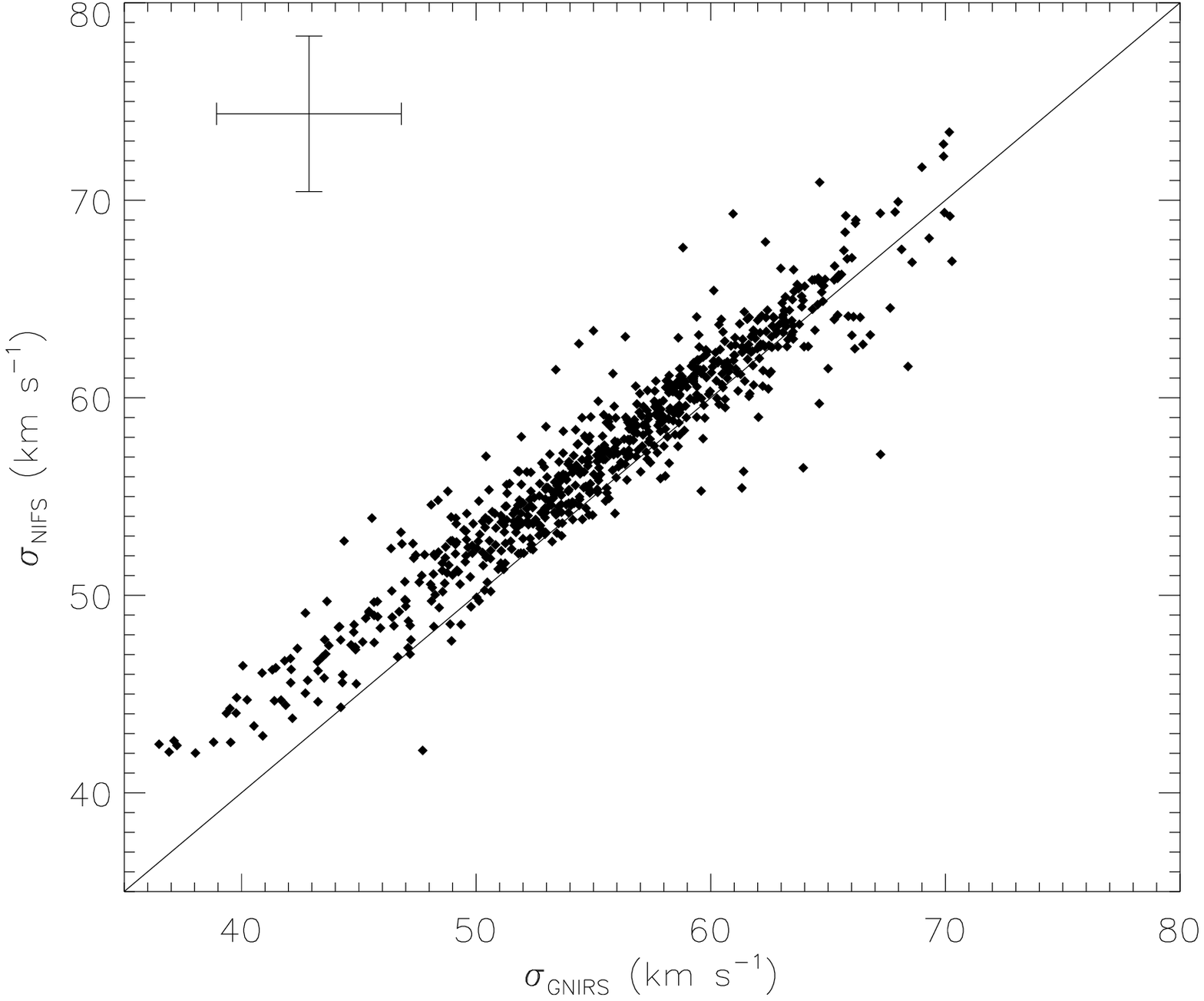}
\caption{Correlation between the velocity dispersion measured from the
NGC4051 datacube using only the stellar templates from the GNIRS
sample [$\sigma$(GNIRS)] and from the NIFS sample
[$\sigma$(NIFS)]. The left panel shows the result using the
templates in their native resolution, while in the right panel, the
resolution of the GNIRS templates was degraded  to match the nominal
resolution of the NIFS grating.  \label{correlation}}
\end{figure}




\begin{deluxetable}{lcccccccc}
\tablecolumns{9}
\tabletypesize{\scriptsize}
\tablecaption{The GNIRS sample.\label{gnirsobs}}
\tablewidth{436pt}
\tablehead{
\colhead{Object} & \colhead{Mag} & \colhead{Sp Type} & \colhead{T$_{\rm eff}$} & \colhead{log g} &
\multicolumn{2}{c}{Blue Setting (2.15-2.33\mic)} & \multicolumn{2}{c}{Red Setting (2.24-2.43\mic)} \\
\colhead{} & \colhead{(V)} & \colhead{} & \colhead{(K)} & \colhead{} & 
\colhead{Date}  & \colhead{Res. (\AA)} & \colhead{Date} & \colhead{Res. (\AA)}}
\startdata
HD 20038   & 8.91 & F7 IIIw & 5196 & 2.62 & 2006 Oct 13 & 2.97 & 2006 Sep 12 & 2.76   \\
           &      &         &      &      & 2006 Oct 18 &      & 2006 Oct 07 &   \\
           &      &         &      &      &             &      & 2006 Oct 13 &   \\
HD 209750  & 2.90 & G2 Ib   & 5091 & 1.45 & \nodata & \nodata  & 2006 Oct 20 & 2.99  \\
HD 6461    & 7.65 & G3 V    & 5110 & 2.30 & 2006 Oct 10 & 2.98 & 2006 Sep 05 & 2.93  \\
           &      &         &      &      &             &      & 2006 Oct 07 &   \\
HD 173764  & 4.23 & G4 IIa  & 4700 & 0.94 & 2006 Oct 12 & 3.02 & 2006 Sep 04 & 2.89   \\
           &      &         &      &      &             &      & 2006 Oct 21 &   \\
HD 36079   & 2.84 & G5 II   & 5170 & 2.27 & 2007 Jan 04 & 3.16 & 2006 Sep 14 & 3.02 \\
HD 1737    & 5.17 & G5 III  & 4790 & 3.05 & 2006 Oct 08 & 2.93 & 2006 Sep 14 & 3.02  \\
HD 213789  & 5.89 & G6 III  & 4820 & 2.88 & \nodata & \nodata & 2006 Sep 04 & 2.89  \\
HD 212320  & 5.92 & G6 III  & 4790 & 2.87 & 2006 Oct 06 & 2.94 & 2006 Sep 04 & 2.89  \\
HD 213009  & 3.97 & G7 III  & 4800 & 2.0  & \nodata & \nodata & 2006 Sep 03 & 3.10  \\
HD 35369   & 4.14 & G8 III  & 4880 & 2.76 & 2006 Oct 15 & 2.96 & 2006 Oct 05 & 2.88  \\
HD 64606   & 7.44 & G8 V    & 5040 & 4.0  & 2007 Jan 04 & 3.16 & 2007 Jan 02 & 2.79  \\
HD 224533  & 4.89 & G9 III  & 4960 & 3.19 & 2007 Jan 07 & 2.79 & 2006 Sep 14 & 3.02  \\
HD 4188    & 4.78 & K0 III  & 4755 & 2.90 & 2006 Oct 12 & 3.02 & 2006 Oct 05 & 2.88  \\
HD 206067  & 5.11 & K0 III  & 4740 & 2.73 & 2006 Oct 06 & 2.94 & 2006 Aug 31 & 2.99  \\
HD 34642   & 4.83 & K0 IV   & 4730 & 3.39 & 2006 Oct 17 & 2.89 & 2006 Sep 12 & 2.71  \\
HD 198700  & 3.66 & K1 II   & 4383 & 0.80 & \nodata & \nodata  & 2006 Sep 04 & 2.89  \\
           &      &         &      &      &         &          & 2006 Oct 21 &   \\
HD 218594  & 3.66 & K1 III  & 4430 & 2.34 & 2006 Oct 20 & 2.93 & 2006 Sep 03 & 3.10  \\
HD 26965   & 4.41 & K1 V(a) & 5091 & 4.31 & 2006 Dec 12 & 3.07 & 2006 Sep 13 & 3.01  \\
           &      &         &      &      &             &      & 2006 Oct 01 &   \\
HD 39425   & 3.12 & K2 III  & 4582 & 2.80 & 2006 Dec 12 & 3.07 & 2006 Oct 01 & 3.01   \\
HD 38392   & 6.15 & K2 V    & 4990 & 4.50 & 2006 Oct 17 & 2.89 & 2006 Oct 01 & 3.01  \\
HD 4730    & 5.61 & K3 III  & 4220 & 2.10 & 2006 Oct 08 & 2.93 & 2006 Sep 03 & 3.10  \\
HD 191408  & 5.31 & K3 V    & 4893 & 4.55 & \nodata & \nodata & 2006 Sep 05 & 3.01  \\
           &      &         &      &      &         &         & 2006 Oct 01 &   \\
HD 9138    & 4.84 & K4 III  & 4040  & 1.91 & 2006 Oct 19 & 2.97 & 2006 Sep 10 & 2.94  \\
HD 720     & 5.42 & K5 III  & 4160 & 2.02 & 2006 Oct 12 & 3.02 & 2006 Sep 10 & 2.94  \\
HD 32440\tablenotemark{a}   & 5.47 & K6 III & \nodata & \nodata & 2007 Jan 06 & 2.79 & 2007 Jan 04  & 2.73  \\
HD 63425B\tablenotemark{a}  & 7.71 & K7 III & \nodata & \nodata & 2006 Dec 12 & 3.07 & 2006 Dec 12 & 2.75  \\	
HD 113538\tablenotemark{a,b}  & 9.02 & K8 V & \nodata & \nodata & 2007 Jan 04 & 3.16 & 2007 Jan 06 & 2.90  \\
HD 2490    & 5.43 & M0 III  & 3652 & 4.0 & 2006 Oct 15 & 2.96 & 2006 Sep 11 & 2.95  \\
HD 112300  & 3.38 & M3 III  & 3652 & 1.3 & \nodata & \nodata & 2007 Jan 16 & 2.89  \\
\enddata
\tablecomments{The spectral resolution in columns (7) and (9) has been
measured from the Ar arc lines. For those stars observed in more than
one night, the value given is the resulting resolution for the combined spectrum}
\tablenotetext{a}{Not in the \citet{strobel97} list; magnitude and
spectral type from Simbad}
\tablenotetext{b}{In Fig.\ref{libew}, this star falls quite off the
overall trend of EW with T$_{eff}$. The spectral type quoted here is
from older references. \citet{gray06} give it a K9V spectral
type (thus moving the point further to the left), but also indicate
presence of (minor) chromospheric activity in this star. The spectrum
is shown in Fig.\ref{gnirsplotlast}.}
\end{deluxetable}


\begin{deluxetable}{lccccc}
\tablecaption{The NIFS sample \label{nifsobs}}
\tablewidth{0pt}
\tablehead{
\colhead{Object}  & \colhead{Mag (V)}  & \colhead{Sp Type}  & \colhead{Date}  & \colhead{$\lambda_c$ (\mic)} & 
\colhead{Resolution (\AA)}
}  
\startdata
HD 210885    & 7.28 & G8 II & 2007 Oct 15  & 2.25  & 3.17 \\
HD 107467    & 7.37 & G8 III & 2006 Jan 18  & 2.20 & 3.20 \\
HD  105028   & 7.37 & K0 III & 2007 Apr 30 &  2.20 & 3.14 \\
BD +44337    &  8.56 & K5 Ib & 2006 Dec 30 & 2.25 & 3.22 \\
HD 10598      &  8.31 & K2 III & 2006 Dec 30 & 2.25 & 3.22 \\
HD 109655    &  7.07 & K5 III &  2006 Jan 01 & 2.20 & 3.12 \\
HD 3989       &  7.31 & K5 II & 2007 Oct 15 & 2.25 & 3.17 \\
HD 30354     &  8.47 & M2 III &  2007 Jan 30 & 2.25 & 3.26 \\
HD 236791   &  8.91 & M3 III & 2007 Jan 02 & 2.25 & 3.23 \\
HD 27796     &  7.75 & M3 III & 2006 Dec 30 & 2.25 & 3.22 \\
HD 235774   &  8.69 & M5 III &  2007 Oct 15 & 2.25 & 3.17 \\
\enddata
\end{deluxetable}


\end{document}